\documentclass[twocolumn,superscriptaddress,aps,pre,showpacs,footinbib]{revtex4}
\pdfoutput=1

\usepackage{graphicx}
\usepackage{amsfonts}
\usepackage{amssymb}
\usepackage{amsmath}

\usepackage{color}

\newcommand{\expect}[1]{\mathbb{E}_{#1}}

\renewcommand{\Re}{{\mathrm{Re}}\, }
\renewcommand{\Im}{{\mathrm{Im}}\, }
\newcommand{\beq}{\begin{equation}}
\newcommand{\eeq}{\end{equation}}
\newcommand{\avg}[1]{\left< #1 \right>}

\newcommand{\barr}{\begin{eqnarray}}
\newcommand{\earr}{\end{eqnarray}}
\newcommand{\Ord}[1]{{\cal O}\left( #1\right)}

\def\Paren#1{\left( #1 \right)}		

\def\figwidth{\columnwidth}

\begin{document}

\title{Griffiths-McCoy singularities, Lee-Yang zeros and the cavity
  method in a solvable diluted ferromagnet}

\author{C. Laumann}

\affiliation{\emph{Department of Physics}, \emph{Joseph Henry
Laboratories},\emph{Princeton University}, Princeton NJ 08544}

\affiliation{\emph{Princeton Center for Theoretical Physics, Princeton
University}, Princeton NJ 08544}

\author{A. Scardicchio}

\affiliation{\emph{Department of Physics}, \emph{Joseph Henry
Laboratories},\emph{Princeton University}, Princeton NJ 08544}

\affiliation{\emph{Princeton Center for Theoretical Physics, Princeton
University}, Princeton NJ 08544}

\affiliation{MECENAS, \emph{Universit\`a Federico II di Napoli}, Via
Mezzocannone 8, I-80134 Napoli, Italy}

\author{S. L. Sondhi}

\affiliation{\emph{Department of Physics}, \emph{Joseph Henry
Laboratories},\emph{Princeton University}, Princeton NJ 08544}

\affiliation{\emph{Princeton Center for Theoretical Physics, Princeton
University}, Princeton NJ 08544}

\begin{abstract} 
  We study the diluted Ising ferromagnet on the Bethe lattice as a
  case study for the application of the cavity method to problems with
  Griffiths-McCoy singularities. Specifically, we are able to make
  much progress at infinite coupling where we compute, from the cavity
  method, the density of Lee-Yang zeroes in the paramagnetic Griffiths
  region as well as the properties of the phase transition to the
  ferromagnet. This phase transition is itself of a Griffiths-McCoy
  character albeit with a power law distribution of cluster sizes.
\end{abstract}


\maketitle

\section{Introduction}

The Bethe-Peierls or cavity method has a long history in statistical
mechanics \cite{baxter82:_exactly_solve}.  The application of this
method to disordered systems has recently undergone a considerable
revival, mainly in connection with the analysis of typical case
complexity of random NP-complete (i.e.\ difficult) optimization
problems. This recent work has led to an improved understanding of the
statistical mechanics of disordered systems---in particular to a
formulation of the physics of replica symmetry breaking without
resorting to replicas. Importantly, it has also led to a new class of
algorithms, now known as survey propagation, for the optimization
problems \cite{Mezard:2002p120}.  In addition, very recent work
\cite{laumann-2007,hastings-2007,Poulin:2007p85} has attempted to
generalize the cavity method/belief propagation to disordered
\emph{quantum} systems, obtaining encouraging results.

The work alluded to above does not address one striking feature of the
physics of disordered systems, namely the presence of Griffiths-McCoy
(GM) singularities
\cite{PhysRevLett.23.17,vojta-2006-39,McCoy:1968p881} in their
thermodynamics in an applied field. While the form of these
singularities can be readily determined from rough estimates of the
statistics of the rare regions from which they emanate, their detailed
extraction can be a tricky task due to their delicate nature,
especially in classical systems. Indeed, in field theoretic
formulations they appear as non-perturbative (instanton) effects
\cite{Dotsenko:1999p2949}.

In this paper we consider the task of extracting these GM
singularities from the cavity method. The method is exact on Bethe
lattices and hence the functional recursion relation to which it gives
rise must contain GM physics which exists already on these
lattices. The challenge then, is to extract it by constructing the
appropriate fixed point solution.  To this end we study the particular
case of a diluted ferromagnet on the Bethe lattice which has an
extended GM region at low temperatures and large dilution. While the
general, exact, determination of GM singularities everywhere in the
phase diagram is a hard problem, we are able to solve the problem in
the infinite coupling limit made precise below. Here we can directly
solve the cavity equations in a field and relate the solution to the
statistics of clusters and to the density of Lee-Yang zeroes commonly
used to characterize GM effects. Further, in this limit the phase
transition between the paramagnet and the ferromagnet is itself
essentially of a GM character and its critical behavior, which we
extract, can be viewed as an enhanced GM phenomenon.

The problem of the dilute Bethe lattice ferromagnet and of GM
singularities has been considered before us
\cite{PhysRevB.40.6980,Barata:1997p4} by different methods. As our
interest is primarily in the development of the cavity method, we give
a self-contained presentation in this paper from that viewpoint. We
turn now to a more detailed enumeration of the contents of this paper.

\section{Model and Organization}

We consider the following disordered Ising Hamiltonian on the Bethe
lattice with connectivity $q$:
\begin{equation} \beta {\cal H}_\epsilon=-J\sum_{\left<ij\right>}
\epsilon_{ij}\sigma_i\sigma_j-H\sum_i\sigma_i,
\end{equation} where
\[ \epsilon_{ij}=\left\{ \begin{array}{ll} 1 & \mbox{with probability
      $p$;}\\ 0 & \mbox{with probability $1-p$}.\end{array} \right. \]
The random couplings $\epsilon_{ij}$ indicate the presence or absence
of a bond in the diluted Bethe lattice. For probability
$p<p_c=1/(q-1)$ the lattice has no giant clusters and the density of
large finite clusters decays exponentially. For $p > p_c$, giant
clusters exist with finite density and at the percolation transition,
$p=p_c$, the density of clusters of size $n$ develops a long algebraic
tail $W_n \sim n^{-5/2}$ (independent of $q$) \cite{fisher:609}. We
note that the dimensionless coupling constants $J$ and $H$ differ from
the conventional magnetic exchange and field by factors of inverse
temperature $\beta=1/T$. The limit $T\to 0$ with $J\gg H$
will be denoted as the $J=\infty$ limit.

In Section \ref{sec:preliminaries}, we provide a guided tour of the
well-known phase diagram of this model from the point of view of the
cavity method. We then establish the critical behavior at the phase
transitions using a set of recursion relations for the moments of the
cavity field distribution. We also show that the critical behavior can
be extracted via a simple numerical algorithm, which we discuss in
some detail in the Appendix.

The following sections are devoted to investigating exact analytic
results in the infinite (dimensionless) spin-spin coupling limit,
$J=\infty$. This corresponds to the horizontal axis of the phase
diagram in Figure \ref{fig:pd}. In Section \ref{sec:cluster}, we find
an explicit expression for the magnetization $M(H)$ by means of a sum
over connected clusters, which follows the standard GM treatment due
to \cite{PhysRevB.12.203}. In Section \ref{sec:cavity} we show that
the same expression can, in fact, be extracted from the cavity
method. In this limit, the magnetization goes to zero with the field
for $p\leq p_c=1/(q-1)$ while for $p>p_c$ a spontaneous magnetization
develops. We first show that: 1) for $p<p_c$ the asymptotic series
expansion for the magnetization contains only integer powers
$M(H)=\chi H+ c_3 H^3+...$ and 2) on the contrary at $p=p_c$ the
series expansion contains semi-integer powers as well
$M(H)=c_{1/2}\sqrt{H}+c_1 H+c_{3/2} H^{3/2}+...$. That is, the
critical exponent $\delta = 2$ at $p = p_c, J=\infty$.

In Section \ref{sec:intrep} we develop an alternative
integral representation for $M(H)$ that corresponds to a harmonic
expansion. This representation will allow us to calculate the (smoothed)
density of Lee-Yang (LY) zeros $\rho_{sm}$ at $J=\infty$ on the imaginary
$H$ axis ($\theta=\Im H$) in Section \ref{sec:leeyang} and to show that
for $p<p_c$ a GM phenomenon indeed occurs, \emph{i.e.} the density
of zeros is non-zero and vanishes as $e^{-\alpha/\theta}$ when
approaching the origin. For $p=p_c$ we find $\alpha=0$ and the
density vanishes as the power law $\rho\propto \sqrt{\theta}$.

Finally, the promised Appendix briefly describes the ``population
dynamics'' algorithm used in the numerical work.

\section{Phase diagram and cavity equations}
\label{sec:preliminaries}

\begin{figure}[tbp]
\begin{center}
\includegraphics[width=\figwidth]{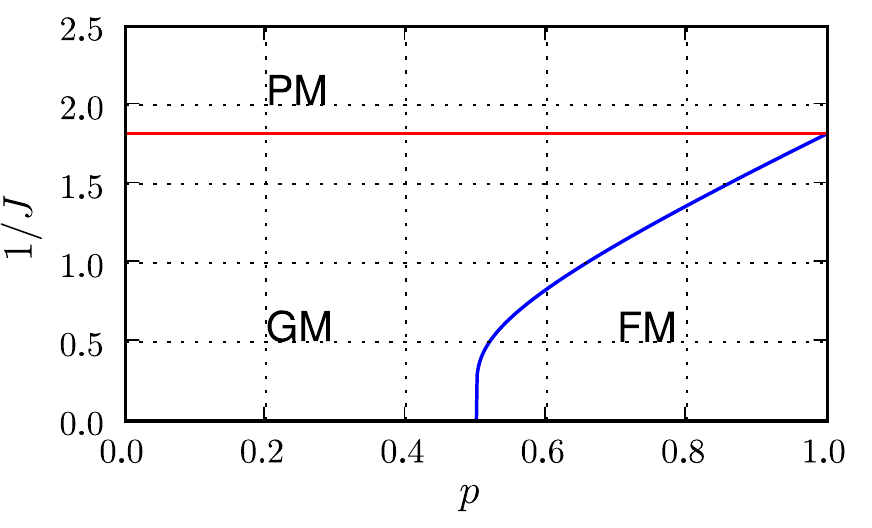}
\caption{Phase diagram for the diluted ferromagnet on the
  connectivity $q=3$ Bethe lattice.}
\label{fig:pd}
\end{center}
\end{figure}

The $p$---$J$ phase diagram (Figure \ref{fig:pd}) of the diluted
ferromagnet ${\cal H}_\epsilon$ is physically well understood and can
be derived naturally in a cavity method formalism
\cite{yedidia00generalized,Mezard:2001p84}. In this approach, one
considers the flow of cavity fields from the boundaries of the tree
inward toward the center. A \emph{cavity field} $h_i$ on a spin
$\sigma_i$ at a distance $d$ from the boundary describes the spin's
magnetization in the absence of the link connecting it to the next
spin inward. The cavity field $h_i$ only depends on the cavity fields
on $\sigma_i$'s neighbors at distance $d-1$ and therefore one can
define a natural flow for the depth dependent distribution of fields
$P^{(d)}(h)$:
\begin{widetext}
  \begin{equation}
    \label{eq:cavdistflow}
    P^{(d)}(h) = 
    \expect{\epsilon}\int \Paren{\prod_{i=1}^{q-1} dh_i P^{(d-1)}(h_i)} 
    \delta\left(h-\sum_{i=1}^{q-1}u(h_i+H, J \epsilon_{i})\right)
  \end{equation}
where
\begin{equation}
  \label{eq:cavbiasdef}
  u(h_i+H,J\epsilon_{i}) = 
  \tanh^{-1}\left(\tanh (J\epsilon_{i})\tanh (h_i+H)\right)
\end{equation}
\end{widetext}
gives the bias on the field $h$ due to a spin $\sigma_i$ connected
through a link $J\epsilon_{i}$. $\expect{\epsilon}$ is the
expectation with respect to the $\epsilon_{ij}$ distribution. Fixed
point distributions $P^{(\infty)}(h)$ describe the statistical
features of the bulk (central region) of the Bethe lattice. In order
to break the Ising symmetry, we will always assume an infinitesimal
uniform positive boundary field $P^{(0)}(h) = \delta(h - 0^+)$ as the
starting point for the flow.

In the undiluted model, $p=1$, all of this discussion reduces to the
simple Bethe-Peierls mean field theory for a connectivity $q$
lattice. Since there is no randomness, the cavity field distributions
$P^{(d)}$ are simply delta functions located at, possibly depth dependent,
fields $h^{(d)}$. Equation (\ref{eq:cavdistflow}) reduces to a flow
equation for $h^{(d)}$:
\begin{eqnarray}
  \label{eq:betheflow}
  h^{(d)} & = & \sum_{i=1}^{q-1}u(h^{(d-1)}+H, J) \\
  & = & (q-1) \tanh^{-1}\Paren{\tanh(J) \tanh(h^{(d-1)}+H)}\nonumber
\end{eqnarray}
For $J < J_G = \tanh^{-1}\left(\frac{1}{q-1}\right)$, the flow at
$H=0$ has only one fixed point $h^{(\infty)}=0$ corresponding to the
paramagnetic phase. For $J > J_G$, the $h^{(\infty)}=0$ fixed point
becomes unstable to a spontaneously magnetized ferromagnetic fixed
point with $h^{(\infty)}>0$. Expansion of the fixed point equation to
leading order in $H$ and $\epsilon=(J-J_G)$ gives the well-known
mean-field critical exponents at $J = J_G$:
\begin{equation}
   \label{eq:bethecritexp}
   \begin{array}{rlllll}
   M(H, J=J_G) & \sim &  H^{1/\delta}; &\mbox{~ }\delta & = & 3 \\
   M(H=0, J > J_G) & \sim & (J-J_G)^{\beta}; &\mbox{~ }\beta & = & 1/2
   \end{array}
\end{equation}

Under dilution, we must return to the more general cavity distribution
flow defined by equation (\ref{eq:cavdistflow}) to extract the phase
behavior. Notice that the paramagnetic cavity distribution $P^{PM}(h)
= \delta(h)$ is always a fixed point of the flow at $H=0$, just like
$h^{(d)}=0$ is always a solution for the undiluted model. As in
undiluted case, this fixed point will become unstable above some
critical coupling $J_c(p)$. Near $P^{PM}(h)$ (\emph{i.e.} for small
$h$), we consider the linear stability of the first moment of
$P^{(d)}(h)$:
\begin{eqnarray*}
  \label{eq:PMlinstab}
  \avg{h}^{(d)} & = & \expect{\epsilon}\int dh 
    \Paren{\prod_{i=1}^{q-1} dh_i P^{(d-1)}(h_i)} 
    \delta\Paren{h - \sum_{i=1}^{q-1} u_i} h\\
  & = & \expect{\epsilon} \int \Paren{\prod_{i=1}^{q-1} dh_i P^{(d-1)}(h_i)}
                               \sum_{i=1}^{q-1} u(h_i, J\epsilon_i) \\
  & \approx & (q-1) \Paren{\expect{\epsilon} \tanh(J\epsilon)} \avg{h}^{(d-1)}
\end{eqnarray*}
to leading order. Thus, $1 = (q-1) p \tanh(J_c(p))$ gives the critical
boundary separating a stable paramagnetic phase from the ferromagnetic
phase. A small rearrangement gives:
\begin{equation}
  \label{eq:phaseboundary}
  J_c(p) = \tanh^{-1}\Paren{\frac{p_c}{p}}
\end{equation}
This agrees precisely with the undiluted critical point $J_c(p=1) =
J_G$ found above and also predicts that for $p < p_c$ the paramagnetic
phase persists for all finite $J$. There is no ferromagnetic
phase transition for a model with only finite clusters, as one expects.

In order to extract the critical behavior along the diluted phase
boundary, we wish to expand the fixed point equations near the
critical solution as we did in the discussion of the undiluted
model. Rather than working with equation (\ref{eq:cavdistflow})
directly, it is more natural to use an equivalent infinite set of
recursion relations for the moments of $P(h)$. These can be derived by
multiplying both sides of equation (\ref{eq:cavdistflow}) by $h^n$ and
integrating or by considering the relation on random variables
\begin{equation*}
  \label{eq:rvflow}
  h'^{(d)} = \sum_{i=1}^{q-1}\epsilon_i\tanh^{-1}\Paren{\tanh(J) \tanh(h_i^{(d-1)})}
\end{equation*}
taken to the power $n$ and averaged. Near $P^{PM}(h)$, we expand this
relation around small $h_i$:
\begin{equation}
  \label{eq:rvflowexp}
  h' = \sum_{i=1}^{q-1}\epsilon_i T (h_i - \frac{1}{3}(1-T^2) h_i^3) + ...
\end{equation}
where $T=\tanh J$ and we have suppressed the depth superscripts.

Sufficiently near the phase boundary, we expect the moments
$\avg{h^n}$ to decrease exponentially with $n$ and thus only a few
leading order moments need be retained to extract the leading critical
behavior at finite $J$. Taking powers of equation (\ref{eq:rvflowexp})
and averaging, we find
\begin{eqnarray}
\avg{h'}&=&2pT\avg{h}-\frac{2}{3}pT(1-T^2)\avg{h^3}\nonumber \\
\avg{h'^2}&=&2pT^2\avg{h^2}+2p^2T^2\avg{h}^2\nonumber \\
\avg{h'^3}&=&2pT^3\avg{h^3}+6p^2T^3\avg{h}\avg{h^2}\nonumber 
\label{eq:hp3}
\end{eqnarray} 
to cubic order. We have specialized to the case $q=3$ in order to
simplify the presentation; for $q>3$ an additional term at cubic order
is generated but the critical exponents remain the same.

Near the phase boundary in the $p$---$J$ plane, we can define a small
parameter $\epsilon$ by writing $2pT=\frac{p}{p_c}\tanh
J=1+\epsilon$. We will treat the fixed point equations to leading
order in the $\epsilon$ expansion. For $\epsilon<0$ the only real
solution is paramagnetic:
\begin{equation}
\avg{h}=\avg{h^2}=\avg{h^3}=0.
\end{equation}
This solution is stable since its local Lyapunov exponents
$(\epsilon,\epsilon-\ln\frac{1}{T},\epsilon-2\ln\frac{1}{T})$ are all negative.

\begin{figure}[tbp]
\centering
\includegraphics[width=\figwidth]{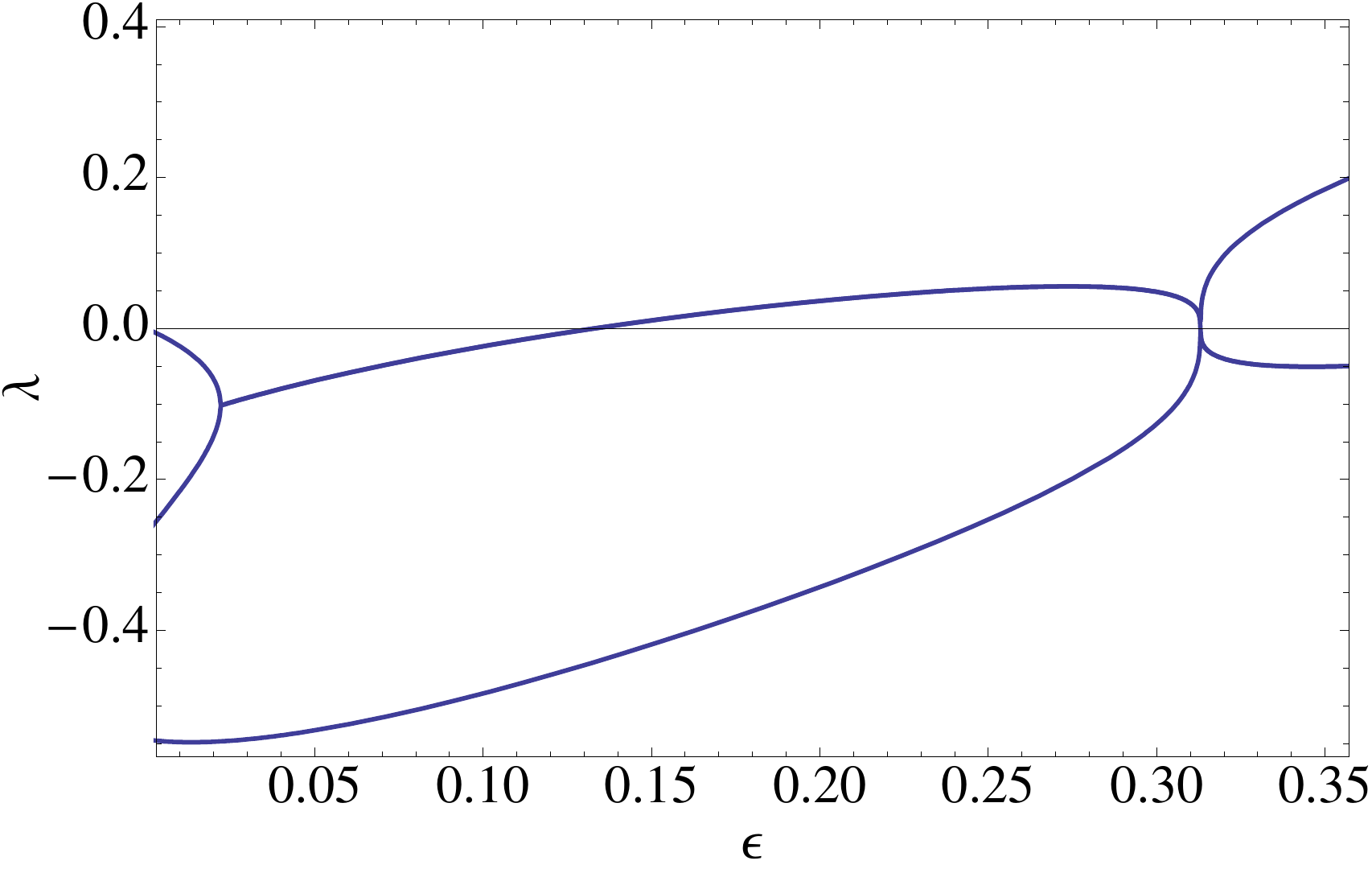}
\caption{Lyapunov exponents of the ferromagnetic fixed point of the
  iteration equations for $J=1$. Notice that for $\epsilon < 0.13...$
  they are all negative, signaling stability of the solution. At
  larger $\epsilon$, the stationary point becomes a focus before
  eventually becoming unstable.}
\label{fig:Lyap}
\end{figure}

For $\epsilon>0$ this solution becomes unstable. We find two other
ferromagnetically ordered solutions, which are linked by the symmetry
$h\to-h$, and choose the positive one. This is
\begin{eqnarray}
\avg{h}&=&2\sqrt{\frac{1-T}{T}}\epsilon^{1/2}+...\ \nonumber\\
\avg{h^2}&=&\frac{2}{T}\epsilon+...\ \nonumber\\
\avg{h^3}&=&\frac{6}{\sqrt{T(1-T)}}\frac{\epsilon^{3/2}}{1+T}+...\ .
\label{eq:ferrostat}
\end{eqnarray}
One can find the Lyapunov exponents of this stationary point
analytically but the expressions are unenlightening. We plot a typical
case in Figure \ref{fig:Lyap}. Notice that the results
(\ref{eq:ferrostat}) are consistent with the assumption that
$\avg{h^n}$ decreases exponentially with $n$.

From Eq.\ (\ref{eq:ferrostat}) one can read off the critical exponent
$\beta=1/2$ as the power of $\epsilon$ in $\avg{h}\sim m$. This is
valid for all $T<1$ and sufficiently small $\epsilon$. The point $T=1$
($J=\infty$) is different and needs to be treated more carefully.  As
$T\to 1$ the coefficient of $\epsilon^{1/2}$ in (\ref{eq:ferrostat})
vanishes, which implies that the $J=\infty$ critical exponent
$\beta'>1/2$ while the divergence of the coefficient of
$\epsilon^{3/2}$ means that $\beta'<3/2$. Indeed, from the exact
solution of Section \ref{sec:cluster}, we will find $\beta'=1$.

For sufficiently large $\epsilon > \epsilon_c$, the Lyapunov exponents
become positive, signaling a loss of stability of the third order
ferromagnetic solution (for the value $J=1,\ T=\tanh(1)$ in Figure
\ref{fig:Lyap}, $\epsilon_c=0.137...\ $). This indicates that the first
few moments flow to large scale and our truncation to cubic order
fails. The value of $\epsilon_c$ decreases monotonically as $T$
approaches $1$ and to accurately find the fixed points we need to keep
track of more moments of $h$ in our iteration equations. At this point
it is convenient to switch to numerical solution of the full cavity
equation (\ref{eq:cavdistflow}) by population dynamics, as described
in the Appendix.

Having explored both above and below the critical point, we return
briefly to the critical point at $\epsilon = 0$. Here, at linear
order, there is a marginal flow near the paramagnetic fixed point. It
is possible to analyze the truncated flow equations (\ref{eq:hp3}) at
higher order to discover that the paramagnetic solution is indeed
algebraically (rather than exponentially) stable, as one expects of a
second order phase transition. With some additional algebra it is
possible to carry a small applied field $H$ through all of the above
arguments at $\epsilon=0$ and show that the critical exponent
$\delta=3$ all along the $p>p_c$ phase boundary.

The final important feature of the phase diagram is the presence of GM
singularities throughout the $p < 1$, $J > J_G$ region. That is, the
density of LY zeros on the imaginary $H$ axis of the partition
function has an essential singularity like $e^{-a'/\Im{H}}$ throughout
this region due to the cumulative influence of rare large undiluted
regions and there is therefore no gap. Equivalently, the real
magnetization $M \sim e^{-a/H}$ in a real applied field $H$. Although
this can be seen from elementary rigorous arguments
\cite{PhysRevLett.23.17,vojta-2006-39}, it is difficult to detect
either analytically or numerically at finite $J$. However in Sections
\ref{sec:intrep} and \ref{sec:leeyang} we will use the exact solution
of the cavity equations at $J=\infty$ to exhibit these essential
singularities explicitly and subject them to detailed study.

\section{Cluster series at $J= \infty$}
\label{sec:cluster}

For the remainder of the paper, we will focus primarily on the
$J=\infty$ part of the phase diagram of the model. We
first review the classic argument due to Harris \cite{PhysRevB.12.203}
based on an expansion over connected clusters. This will lead to an
exact series expansion for $M(H)$ that we will
independently rederive using the cavity approach in Section
\ref{sec:cavity}.

Consider a cluster of $n+1$ spins connected by $n$ bonds. For $J\gg
H\sim 1$ (which is the meaning of the $J=\infty$ limit) each connected
cluster behaves like a piece of ferromagnet. Indeed for $H=0$ there
are two degenerate ground states, one with all spins pointing up and
one with all spins pointing down. The first excited states are spin
flips at energy $\sim J$ above the ground states and their presence is
negligible. Turning on a magnetic field $H$ the degeneracy is broken
and (if $H$ is positive, say) the state with all spins pointing up is
energetically preferred. Therefore the cluster will acquire a small
magnetization:
\begin{equation}
M_n(H)=(n+1) \tanh ((n+1) H).
\end{equation}
The total magnetization per spin is obtained by summing over all the
clusters with their weights $W_n$, corresponding to the number of
clusters of size $n$ per spin \footnote{Notice that in this approach
  the infinite coupling limit is solvable because the magnetization of
  a cluster depends only on the number of spins in the cluster and not
  on its detailed shape.}
\begin{equation}
\label{eq:basicM}
M(H)=\sum_{n\geq 0} W_n M_n(H).
\end{equation}
This equation has been studied before and results can be
found in \cite{PhysRevB.12.203} for the magnetization, and
\cite{0022-3719-9-2-022} for the scaling law of the magnetization at
the critical point. We will reproduce those results on the
magnetization for completeness, but the main focus of this paper will
be the density of Lee-Yang zeros and the solution of cavity field
equations from which we will recover the known results.

From the solution of the bond percolation problem on the Bethe lattice
\cite{fisher:609} the number per spin $W_n$ of clusters of bond size
$n$ is given by
\begin{equation*}
W_n(p)=q\frac{((n+1)(q-1))!}{(n+1)!(n(q-2)+q)!}p^n(1-p)^{n(q-2)+q}.
\end{equation*}
For simplicity we consider $q=3$. Then $p_c=1/2$. We can easily obtain
the asymptotic behavior of the magnetization by using the asymptotics
of $W_n$ as
\begin{equation}
\label{eq:asympt1}
W_n=\frac{12}{\sqrt{\pi}}(1-p)^3\left(\frac{1}{n}\right)^{5/2}e^{-n A(p)}
\end{equation}
where 
\begin{equation}
\label{eq:Ap}
A(p)=\ln\frac{1}{4p(1-p)}.
\end{equation}
$A(p)$ is the exponent governing the decay rate of the cluster sizes
and it will appear often in the remainder of the paper.  For
$p<p_c=1/2$, $A>0$ and $W_n$ decreases exponentially. For $p=p_c=1/2$,
$A=0$ and we have instead a power-law decay with exponent $5/2$ (the
exponent is independent of $q$):
\begin{equation}
W_n=\frac{3}{2\sqrt{\pi}}\left(\frac{1}{n}\right)^{5/2}.
\end{equation}
This change in the asymptotic fall-off of the cluster distribution at
criticality is the reason for the change in the response to an applied
external field at zero temperature.

Indeed we can easily see how this works. For $A>0$ and small $H$ we
can write an asymptotic expansion:
\begin{eqnarray}
\label{eq:clustermh}
M(H)&=&\sum_n W_n (n+1) \tanh ((n+1) H)\nonumber\\
&\simeq& H \avg{(n+1)^2}+\Ord{H^3}\ \nonumber\\
&=&H\frac{(1+p)}{1-(q-1)p}+\Ord{H^3},
\end{eqnarray}
which is linear in $H$. \emph{At} the percolation threshold, however,
$A=0$ and $\avg{n^2}$ diverges. The expansion of $\tanh$ inside the
first sum is unjustified. To find the first term in the asymptotic
expansion of $M$ (that we will derive in a formally correct way in
Section \ref{sec:intrep}) we use instead (\ref{eq:asympt1}):
\begin{eqnarray}
\label{eq:heuristM}
M(H)&\simeq&\sum_n \frac{3}{2\sqrt{\pi}}\left(\frac{1}{n}\right)^{3/2}\tanh( n H) \\
&\simeq& \frac{3}{2\sqrt{\pi}}\sqrt{H}\int_0^\infty dx\ x^{-\frac{3}{2}}\tanh x+\Ord{H}. \nonumber
\end{eqnarray}
So the susceptibility diverges although there is no spontaneous
magnetization \cite{PhysRevB.12.203}. We now turn to a derivation of
the above results from the cavity method.

\section{Cavity approach at $J = \infty$}
\label{sec:cavity}

The cavity method for this system gives a probability distribution for
the cavity fields which satisfies the fixed point equation
(cf. Equation (\ref{eq:cavdistflow}))
\begin{equation}
\label{eq:cav-consistency}
P(h)= \expect{\epsilon}\int \prod_{i=1}^{q-1} dh_i P(h_i)\delta\left(h-\sum_{i=1}^{q-1}u(h_i+H, J \epsilon_i)\right)
\end{equation}
In the $J\to \infty$ limit, we can linearize the cavity biases $u$: 
\begin{equation}
\label{eq:cav-linear}
u(h+H,J \epsilon_i)=(h+H)\epsilon_i
\end{equation}
At $q=3$, the fixed-point equation (\ref{eq:cav-consistency}) becomes
\begin{eqnarray*}
  P(h)&=&(1-p)^2\delta(h)+2p(1-p)P(h-H)\nonumber\\*
  &+&p^2\int dh_2 P(h_2-2 H)P(h-h_2).
\end{eqnarray*}
This equation can be solved by defining the Laplace transform
\begin{equation*}
g(s)=\int_{0^-}^\infty dh P(h)e^{-s h},
\end{equation*}
making sure to include the delta function at $h=0$. The equation for
$g$ is quadratic
\begin{eqnarray}
\label{eq:eqforgs}
0 & = & p^2e^{-2sH}g(s)^2+(2p(1-p)e^{-sH}-1)g(s)\nonumber\\*
&+&(1-p)^2.
\end{eqnarray}
with solution
\begin{equation}
\label{eq:solgs}
g(s)=\frac{e^{2 H s}-2 e^{H s}(p- p^2)-e^{\frac{3 H s}{2}} \sqrt{e^{H s}-4(p-p^2)}}{(2 p^2)}.
\end{equation}
The second solution to (\ref{eq:eqforgs}) is not physical. Even
without inverting the Laplace transform all the properties of the
solution can be extracted from $g(s)$. For example the normalization
condition, the zeroth moment, is
\begin{equation*}
\int_0^\infty dh P(h)= g(0)=\left\{ \begin{array}{ll}
         1 & \mbox{if $p\leq 1/2$;}\\
         \frac{(1-p)^2}{p^2} & \mbox{if $p>1/2$}.\end{array} \right.
\end{equation*}
and the first moment is
\begin{equation}
\label{eq:pfirstmom}
\avg{h}=-\frac{\partial g}{\partial s}\Big|_{s=0}=\frac{2Hp}{1-2p}
\end{equation}
whose divergence at $p=p_c=1/2$ signals the ferromagnetic phase
transition.  For $p>p_c$, $P(h)$ loses normalization because a finite
fraction of the cavity fields flow to infinity, just as in a
percolating cluster distribution. Indeed, the divergent cavity fields
are precisely those attached to spins in percolating clusters. Because
these spins are connected to the positively biased boundary and the
temperature is effectively zero, they spontaneously magnetize to
$M=1=\tanh(\infty)$. This provides the spontaneous magnetization
critical exponent:
\begin{equation*}
M(H=0,J=\infty,p)= 1-\frac{(1-p)^2}{p^2}\propto (p-p_c)^1
\end{equation*}
from which we read $\beta'=1$ in accord with the discussion following
Eq.\ (\ref{eq:ferrostat}).  

We now concentrate on the $p\leq p_c=1/2$ region at finite $H$. Consider
the magnetization per spin
\begin{equation}
\label{eq:cavmag}
M(H) = \avg{\tanh(H+\sum_{i=1}^q u(h_i+H,J\epsilon_i))}_{\epsilon,h_i}
\end{equation}
which is obtained by averaging over the disorder and the distribution
of $h$. It is straightforward to show that for small $h$ and $H$ we obtain
the results of Equation (\ref{eq:clustermh}). Indeed:
\begin{eqnarray}
M(H) &= &\avg{H+\sum_{i=1}^q\epsilon_i(h_i+H)}_{\epsilon,h}+\Ord{H^3}\\
&=&(1+pq)H+qp\avg{h}+\Ord{H^3};
\end{eqnarray}
Now substitute (\ref{eq:pfirstmom}) and simplify
\begin{equation}
M(H)=H\frac{(1+p)}{1-(q-1)p}+\Ord{H^3},
\end{equation}
in accordance with (\ref{eq:clustermh}).

We now reconstruct the full probability distribution $P(h)$
exactly. We expand the function $g(s)$ as a series in $e^{-s H}$
\begin{equation}
g(s)=\sum_{n\geq 0} \alpha_n e^{-s n H}
\end{equation}
which defines the coefficients $\alpha_n$. $P(h)$ is now given by the
inverse Laplace transform,
\begin{equation}
\label{eq:Psol}
P(h)=\sum_{n\geq 0}\alpha_n\delta(h-n H).
\end{equation}
The $\alpha_n$ are given by the series expansion of the square root in
(\ref{eq:solgs}):
\begin{equation}
\label{eq:asol}
\alpha_n=\frac{(4p(1-p))^{n+2}}{2p^2}\frac{(-1)^{n+1}\Gamma(3/2)}{\Gamma(n+3)\Gamma(-n-1/2)}.
\end{equation}
That is, $P(h)$ is a comb of delta functions at integer multiples of
$H$ with a decaying envelope. The large $n$ behavior of the envelope
is
\begin{equation}
\label{eq:asollgn}
\alpha_n=\frac{4(1-p)^2}{\sqrt{\pi}}n^{-3/2}e^{-A(p)n}+...\ ,
\end{equation}
where $A(p)$ is defined in (\ref{eq:Ap}). It is not surprising that
the same asymptotics governs both $P(h)$ and $W_n$.

Finally, to connect directly with the previous section let us compute the
exact magnetization of a spin as a function of applied field $H$. Evaluating
the cavity magnetization equation (\ref{eq:cavmag}) using the cavity
field distribution (\ref{eq:Psol}), we find
\begin{widetext}
  \begin{equation}
    M(H) = \sum_{j=0}^{q} \binom{q}{j} p^j (1-p)^{q-j} \sum_{m_1,\cdots,m_j=0}^{\infty}\alpha_{m_1}\cdots\alpha_{m_j}  \tanh{\Paren{(j+1 + m_1+\cdots+m_j) H}} 
  \end{equation}
\end{widetext}
which naturally expands as a series in $\tanh{n H}$. At $q=3$, we can
evaluate all the coefficients in this series to find that indeed they
are identical to the coefficients $n W_{n-1} $ of
Eq. (\ref{eq:basicM}). Thus, at $J =\infty$ the cluster series and the
cavity method produce identical results for the magnetization.

Having established the equivalence of the two solutions, we now return
to the analysis of the series for the magnetization. In the following
two sections we will extract the critical behavior near $p=p_c$ and,
by analytic continuation to imaginary $H$, the GM singularity in the
density of LY zeros.

\section{An Integral representation for the magnetization}
\label{sec:intrep}

Despite its simplicity, the expansion of Eq.\ (\ref{eq:basicM}) is
an exact result for the magnetization which can be analytically
continued to imaginary values of the magnetic field. However, the
representation of $M(H)$ as a sum in (\ref{eq:basicM}) is not best
suited for this purpose. An integral representation would be
preferable. To obtain it we write the Laplace transform of the
function $\tanh x$
\begin{eqnarray}
f(s)&=&\int_0^\infty dx e^{-sx}\tanh x \nonumber\\
&=&\frac{1}{2} \left(-\frac{2}{s}-\psi\left(\frac{s}{4}\right)+\psi\left(\frac{2+s}{4}\right)\right),
\end{eqnarray}
where $\psi$ is the digamma function. The function $f(s)$ has simples
poles only at the negative even integers and thus we can invert the
transform and write
\begin{equation}
\tanh x=\int_B \frac{ds}{2\pi i}e^{sx}f(s),
\end{equation}
where $B$ is any Bromwich path lying to the right of all poles of
$f(s)$, that is to the right of the negative real axis. Inserting into
(\ref{eq:basicM}), we can invert sum and integral, provided
\begin{equation}
\left|4p(1-p)e^{s H}\right| <  1.
\end{equation}
The resulting expression, valid for $\left| \arg H\right| < \pi / 2$,
\begin{eqnarray*}
M(H)&=&3(1-p)^3\int_B\frac{ds}{2\pi i}f(s)\\
&\times&\sum_{n\geq 0} \frac{2(n+1)!}{(n+3)!n!} e^{s (n+1) H}(p(1-p))^n
\end{eqnarray*}
can be written in closed form by performing the sum. This amounts to
calculating the derivative of the generating function of the
probability $W_n$. For the Bethe lattice the generating function
is \footnote{The generating function can be derived for $q=4$ as well
  while for generic $q$ the result is written in terms of
  hypergeometric functions for which no explicit rational form seems
  to exist. We see below that the fundamental property of this
  generating function is to have a square-root branch cut at
  $\xi=1$. This exists for all $q>2$ due to a well-known property of
  the generalized hypergeometric functions.}
\begin{widetext}
  \begin{eqnarray}
    \phi(x)&=&\sum_{n\geq 0} W_n x^n\nonumber\\
    &=&-2(1-p)^3x^{-3}(8(1-\sqrt{1-\xi})
    + 4(2\sqrt{1-\xi}-3)\xi+3\xi^2)
  \end{eqnarray}
  where $\xi=4p(1-p)x$ has been defined for convenience. By means of
  this function we can perform the sum inside the integral to obtain

  \begin{eqnarray}
    M(H)&=&\frac{3(1-p)^3}{6p^3(1-p)^3}\int_B\frac{ds}{2i\pi}f(s)e^{-2sH}\nonumber\\
    &\times&\left((p(1-p)e^{sH}-1)\sqrt{1-4p(1-p)e^{sH}}+1-3p(1-p)e^{sH}\right).
  \end{eqnarray}
\end{widetext}

We simplify this expression by considering the analytic structure of the
integrand. The function $f(s)$ has simple poles only at the negative
even integers (0,-2,-4,...), while the rational expression has a
series of square root cuts at $s^*_n=\frac{A(p)}{H}+i2\pi n/H$. We
close the contour with a semicircle at infinity on the right (for $\Re
s>0$) on the first Riemann sheet. We then deform the contour to
coincide with the edges of the cuts. At this point only the
discontinuity across the cuts contributes to the final result. For
aesthetic reasons we finally shift the value of $s$ by
$\frac{A(p)}{H}$, the real part of the origins of the cuts.

The resulting expression is
\begin{widetext}
  \begin{equation}
    \label{eq:integrM}
    M(H)=\sum_{n=-\infty}^{\infty}\frac{8(1-p)^2}{\pi p}\int_0^\infty ds\left(1-\frac{1}{4}e^{sH}\right)\sqrt{e^{sH}-1}f(s+s^*_n)e^{-2sH}.
  \end{equation}
\end{widetext}
At this point it seems we have traded a sum of functions
(\ref{eq:heuristM}) with a series of integrals that we cannot
evaluate. This looks like a step backward in the quest for a useful
result! However, after thinking about the procedure we have performed,
we recognize that this is a Poisson summation-like duality on the
original equation (\ref{eq:basicM}). The terms in the sum are higher
and higher harmonics of the result (this is particularly evident, as
we will see shortly, for imaginary $H$).

The series in $n$ in (\ref{eq:integrM}) is dual to the series in
(\ref{eq:basicM}) so that when the first converges rapidly the second
does not and vice versa (for $H$ on the real axis). In the
interesting regime, close to the percolation threshold
(\ref{eq:basicM}) converges slowly and the first term ($n=0$) of
(\ref{eq:integrM}) gives the leading term in the expansion in
$(p-p_c)$ and $H\to 0$.

Let us now see how we can recover Eq.\ (\ref{eq:heuristM}) in a clean
way. At the critical point $p=1/2$, we have $A=0$ and so $s^*_n=i2\pi
n/H$. For $H\to 0$ all the cuts except that corresponding to $n=0$
go to infinity and we can keep only the $n=0$ term in the series
(\ref{eq:integrM}). Moreover, by expanding the integrand in powers of
$H$ we find
\begin{eqnarray*}
M(H)&\simeq&\frac{3}{\pi}\sqrt{H}\int_0^\infty\sqrt{s}f(s)+\Ord{H}\nonumber\\
&=&\frac{3}{\pi}\sqrt{H}\frac{\sqrt{\pi}}{2}\int_0^\infty x^{-3/2}\tanh x+\Ord{H}
\end{eqnarray*}
which coincides with Eq.\ (\ref{eq:heuristM}).

To recapitulate, the magnetization is given by an integral of the
\emph{discontinuous part} of the generating function $\phi$ of the
cluster distribution $W_n$ with the Laplace transform of the function
$\tanh x$. For the Bethe lattice the generating function can be
written down explicitly and the calculations can be carried to the
end. In the percolation limit the cut on the real axis gives the
greatest contribution to the sum.

\section{Density of Lee-Yang zeros at $J=\infty$}
\label{sec:leeyang}

In this Section we will find the density of Lee-Yang zeros $\rho$ at
$J=\infty$. These are the zeros of the partition function as a
function of the external magnetic field $H$, for imaginary
$H=i\theta$. Instead of solving the equation $Z(H)=0$ directly, we
rely on the relation
\begin{equation*}
\rho(\theta)=\frac{1}{\pi}\Re M(i\theta+0^+).
\end{equation*}
To get an idea of how this function looks we recall
\begin{equation*}
\Re\tanh(i\theta+0^+)=\pi\sum_{m=-\infty}^{\infty}\delta(\theta-\frac{\pi}{2}(2m+1)).
\end{equation*}
From this we find
\begin{eqnarray}
\label{eq:rhodelta}
\rho(\theta)&=&\sum_m\sum_{n\geq 0}W_n(n+1)\delta((n+1)\theta-\frac{\pi}{2}(2m+1))\nonumber\\
&=&\sum_{m,n}W_n\delta\left(\theta-\pi\frac{2m+1}{2n+2}\right)
\end{eqnarray}
so the zeros are located at all the
$\frac{\textrm{odd}}{\textrm{even}}$ rational multiples of $\pi$, with
multiplicities given by the $W_n$'s \footnote{Usually, disorder
  averaging generates a smooth density of LY zeros but not so in the
  infinite coupling limit of the diluted ferromagnet. This is because
  the thermodynamics only depend on the size of clusters and not their
  shape and each cluster of size $n$ contributes zeros precisely at
  odd multiples of $\pi/2n$.}.  This is a singular distribution with an
accumulation point at $\theta=0$: our task is now to smooth it by
using the Poisson-dual integral representation (\ref{eq:integrM})
obtained in the previous section.

\begin{figure}[tbp]
\centering
\includegraphics[width=\figwidth]{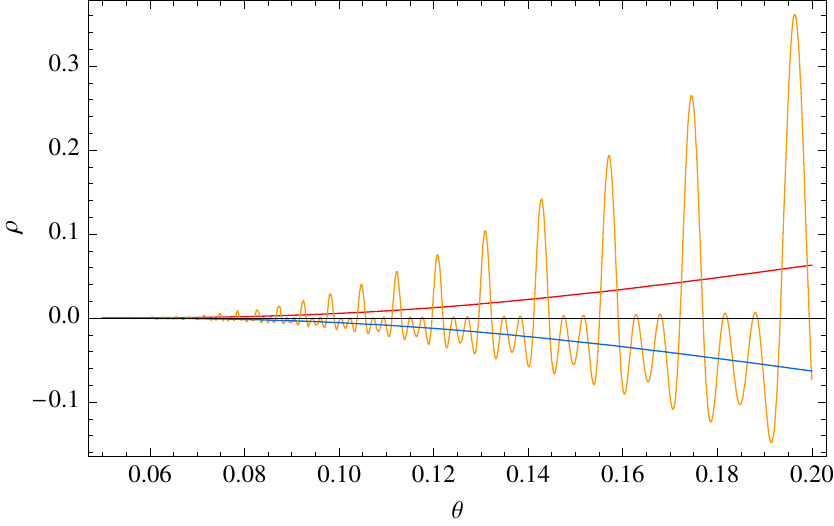}
\caption{The density of Lee-Yang zeros as a function of the imaginary
  field $\theta$ at $p=1/4$. In red the smoothed $\rho_{sm}$, in
  orange the sum of the first three harmonics in (\ref{eq:integrM})
  (terms $n=\pm 1,\pm 2, \pm 3$) and in blue $-\rho_{sm}$. The figure
  suggests (in agreement with the discussion in the text) that the sum
  of all the harmonics (with $n\neq 0$) builds a sum of delta
  functions Eq.\ (\ref{eq:rhodelta}) \emph{minus} $\rho_{sm}$ in
  (\ref{eq:smrho}).}
\label{fig:3harms}
\end{figure}

The expansion over the cuts, for imaginary $H$ becomes an expansion in
higher and higher harmonics (see Figure \ref{fig:3harms}) of
$\rho$. Selecting the term with $n=0$ in (\ref{eq:integrM}),
gives the function smoothed to the lowest degree:
\begin{widetext}
  \begin{equation}
    \rho_{{\mathrm{sm}}}(\theta)=\frac{8(1-p)^2}{\pi^2p}\int_0^\infty ds  \sqrt{e^{s\theta}-1}\left(1-\frac{1}{4}e^{s\theta}\right)e^{-2s\theta}\Im f(-is-iA(p)/\theta+0^+).
  \end{equation}
  This expression simplifies since from the definition of $f$
  \begin{equation}
    \Im f(-i z+0^+)=\int_0^\infty dx \sin (zx)\tanh x =\frac{\pi}{2\sinh\pi z/2}.
  \end{equation}
  So we find the smoothed density of LY zeros
  \begin{equation}
    \label{eq:smrho}
    \rho_{\mathrm{sm}}(\theta)=\frac{4(1-p)^2}{\pi p}\int_0^\infty ds \sqrt{e^{s\theta}-1}\left(1-\frac{1}{4}e^{s\theta}\right)\frac{e^{-2s\theta}}{\sinh\left(\frac{\pi}{2} (s+A/\theta)\right)}. 
  \end{equation}
\end{widetext}

\begin{figure}[tbp]
\centering
\includegraphics[width=\figwidth]{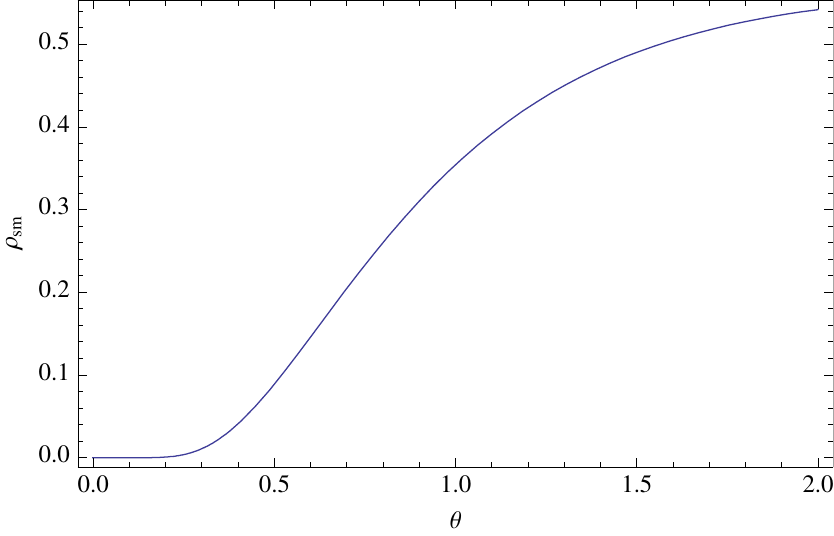}
\includegraphics[width=\figwidth]{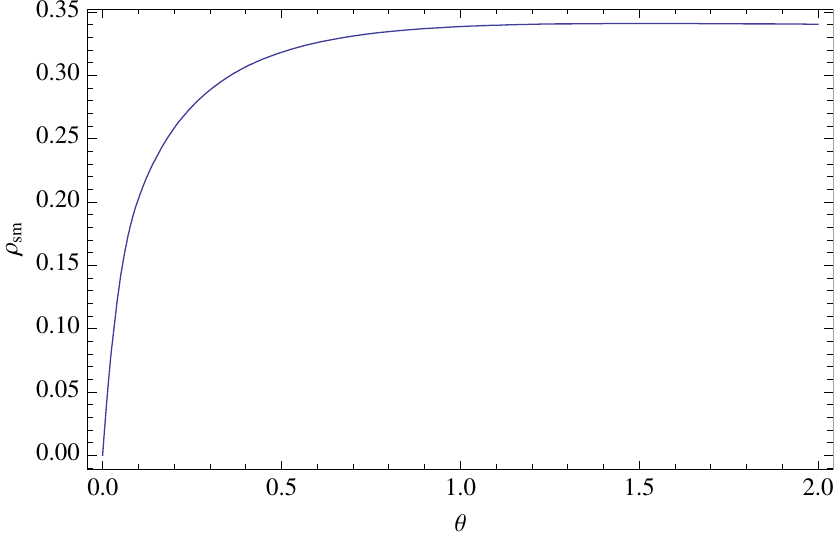}
\caption{The density of Lee-Yang zeros as a function of the imaginary field $\theta$ at $p=1/10$ (above) and at criticality $p=1/2$ (below).}
\label{fig:LYzeros}
\end{figure}

The different profiles for this density can be seen in Figure \ref{fig:LYzeros}.
Here the GM phenomenon is evident: even at $p<p_c=1/2$,
$\rho_{\mathrm{sm}}(\theta)$ is strictly positive for any non-zero
$\theta$; there is no gap in the distribution. This effect is due to
the presence of rare large clusters. The asymptotic expansion of the
density at small $\theta$ can be found by expanding the integrand
\begin{equation}
\rho(\theta)\simeq\frac{3\sqrt{2}(1-p)^2}{4\pi^2 p}\sqrt{\theta}e^{-\frac{\pi}{2}A/\theta}.
\end{equation}
This expansion is uniformly valid at the point $A=0$, which is
$p=p_c$, where it shows the critical square root cusp in the
magnetization. However, let us remark that Eq.\ (\ref{eq:smrho}) is
the smoothed part (in the sense of distributions) for all values of
$\theta$ and not only for small $\theta$.

Let us now make a few qualitative remarks on the asymptotic expansions
for $M(H)$ and $\rho$ which apply in principle to all lattices. From
the integral representation we observe that there is no Stokes
phenomenon for $\Re H>0$, \emph{i.e.}\ the asymptotic approximation
for $M(H)$,
\begin{equation}
\label{eq:asymptM}
M_L(H)= \sum_{k=0}^L a_k H^{2k+1}\avg{(n+1)^{2k+2}}+\Ord{H^{2L+1}}
\end{equation}
(where $a_k$'s are the coefficients in the series expansion for $\tanh
x$) is valid for all $\left|\arg H\right|<\pi/2$. Naively,
substituting $H=i\theta + 0^+$ into this expansion, we obtain a purely
imaginary result for any $L$ and we might speculate that
$\rho=0$. However, the expansion (\ref{eq:asymptM}) is only
asymptotic, since $\avg{(n+1)^k}\sim k!e^{-k A}$. This means that we
cannot take $L\to\infty$ but rather truncate the series at the $L\sim
H/A$ where the remainder is smallest. The remainder is never actually
zero but it is exponentially small in $1/|H|$. A good quantitave
approximation can be found by using the ``terminant''
\cite{dingle73:_asymp_expan} of the asymptotic expansion. The
terminant is indeed $\propto e^{-\frac{\pi}{2}A/|H|}$ and \emph{it is
  not} purely imaginary for $H=i\theta+0^+$. Thus, as a general rule we
expect the real part of the terminant of the asymptotic expansion of
$M(H)$ represents the density of LY zeros in the subcritical region.

\section{Summary}

The diluted Ising ferromagnet on a Bethe lattice is a tractable model
that beautifully illustrates many of the key physical features of
short-ranged disordered systems. In this paper, we have attempted to
present a unified analysis of the model in the framework of the cavity
method, from which we derive both well-known elementary results about
its phases and non-trivial features such as GM singularities and the
infinite coupling critical exponents. 

In particular, the ferromagnetic phase boundary lies in the mean-field
universality class ($\delta = 3$) at any dilution above the
percolation threshold. At this threshold however, the ferromagnetic
critical coupling diverges ($J\to\infty$) and our closed form
solutions for the cavity distributions in this limit reveal that the
critical behavior is governed by the percolation of the underlying
lattice ($\delta = 2$). Linear stability analysis of the flow of the
cavity moments near criticality naturally reveals the Lyapunov
exponents and the associated correlation depth of the stable phases.

Furthermore, at infinite coupling we have explicitly exhibited the
essential Griffiths-McCoy singularities in the magnetization for all
$p<p_c$, where there is no spontaneous magnetization. By an harmonic
resummation of the exact magnetization, we found the smoothed density
of LY zeros exactly and conjectured its relation to the real part of an
appropriate terminant of the asymptotic series for the magnetization. 

\begin{acknowledgments}
  The authors would like to thank M.~Aizenmann for discussions and for
  catching an error in an earlier version of the draft. C.~L.\ and
  A.~S.\ would like to thank S.~Franz and R.~Zecchina for discussions
  and A.~S.\ would like to thank G.~Marmo for discussions and
  acknowledges support from the MECENAS program of the Universita'
  Federico II di Napoli, where part of this work was
  completed. C.~L. acknowledges support from the NSF. The work of
  S.L.S. is supported by the NSF grant number DMR 0213706.
\end{acknowledgments}

\appendix*

\section{Numerical Methods}
\label{sec:numerics}

In the cavity framework, all of the statistical observables of a model
can be derived from the cavity field distribution $P(h)$. This
distribution is the fixed point of the flow of the cavity equation
(\ref{eq:cavdistflow}). While we can solve this equation analytically
in certain limits, we rely on a simple iterative numerical algorithm
called \emph{population dynamics} for many of the finite coupling
results. Population dynamics and its more sophisticated variants
appear in, for example, \cite{mezard-2003-111}.

The algorithm works as follows: we represent the distribution $P(h)$
by a finite population of $N_{pop}$ fields $h_i$. This population is
initialized from an appropriate uniform distribution and then iterated
as follows:
\begin{enumerate}
\item Select $q-1$ fields $h_i$ randomly from the population and
  $q-1$ random $\epsilon_{i0}$.
\item Use (\ref{eq:cavdistflow}) to calculate the cavity field $h_0$
  on a spin sitting below the $q-1$ spins selected above.
\item Randomly replace one element of the population with $h_0$.
\item Repeat until convergence in some measure of the population, for
  example the cavity magnetization $\avg{\tanh(h)}$.
\end{enumerate}
In practice, this procedure converges quickly deep in either the
ferromagnetic or paramagnetic phase but slows near the phase
transition. We illustrate some typical results below for $q=3$.

\begin{figure}
  \centering
  \includegraphics[width=\figwidth]{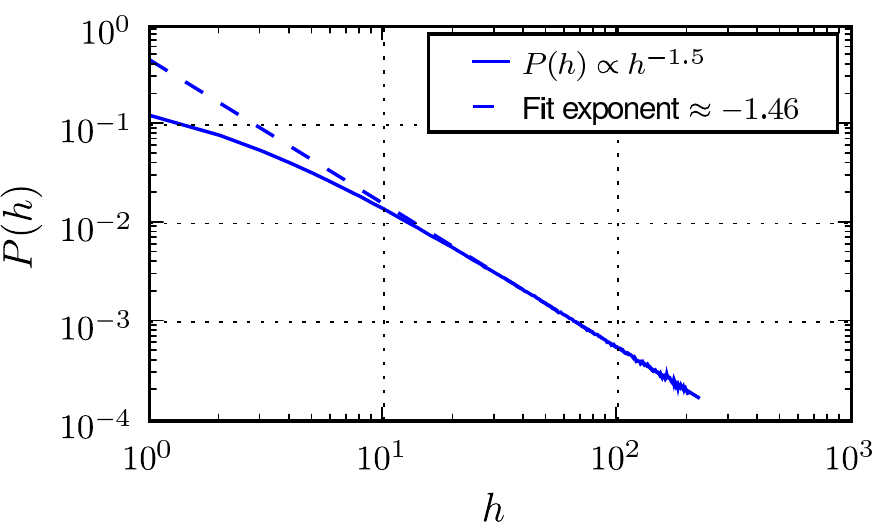}
  \caption{Cavity field distribution at $p=0.5, J=\infty$ critical
    point with $H=1$. This log-log plot shows agreement with the asymptotic form
    $P(h) \sim h^{-3/2}$ found in the exact solution. }
  \label{fig:cavfields-pt2-bz1}
\end{figure}

Even at the percolating critical point, when the expected cavity
distribution develops a long tail and divergent moments, this
procedure works. Figure \ref{fig:cavfields-pt2-bz1} shows the
numerically determined cavity field distribution for the $p=0.5,
J=\infty$ critical point with applied field $H=1$. As noted in
Section \ref{sec:cavity}, the exact solution is a comb of delta
functions at $h = n H, n\in\mathbf{N}$ with weights decaying
asymptotically as a power law $\alpha_n \sim n^{-3/2}$. The
numerical solution concentrates on integer fields with a power-law
tail consistent with the exponent $-1.5$.
 
\begin{figure}
  \centering
  \includegraphics[width=\figwidth]{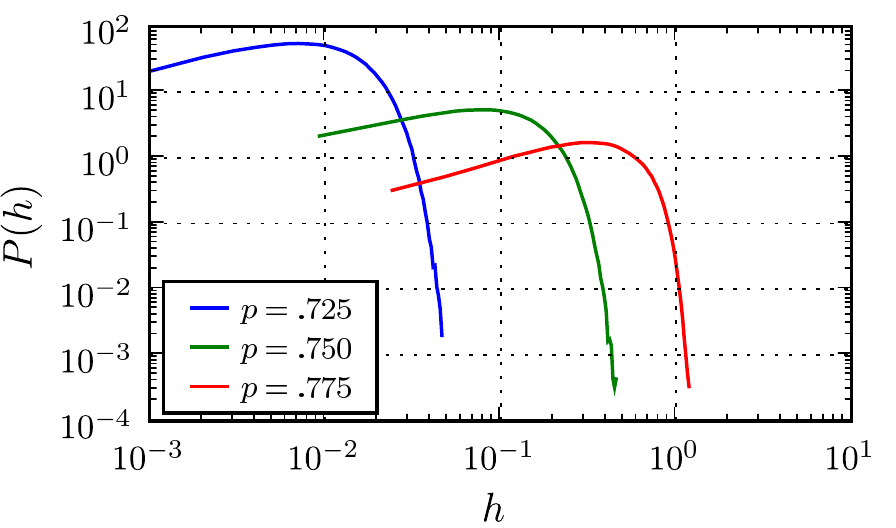}
  \caption{Cavity field distributions as function of $p$ near the
    $p=0.75, J=0.80$ critical point at small applied field $H\approx
    10^{-4}$. The blue curve is in the paramagnetic regime, the green
    is in the critical phase and the red in the ferromagnetic
    one. Notice the dramatic increase in the response of the cavity
    field distribution to the applied field on the ferromagnetic side
    of the phase boundary.}
\label{fig:psweep}
\end{figure}

In general, the form of the cavity field distribution at finite $p$
and $J$ is only obtainable numerically. Figure \ref{fig:psweep} shows
the numerically determined distribution at small applied field $H$ on
three different points in the $p-J$ plane near the $p=0.75, J=0.80$ critical point. These distributions are typical and illustrate the
dramatic increase in susceptibility on the ferromagnetic side of the
phase boundary.

\begin{figure}
  \centering
  \includegraphics[width=\figwidth]{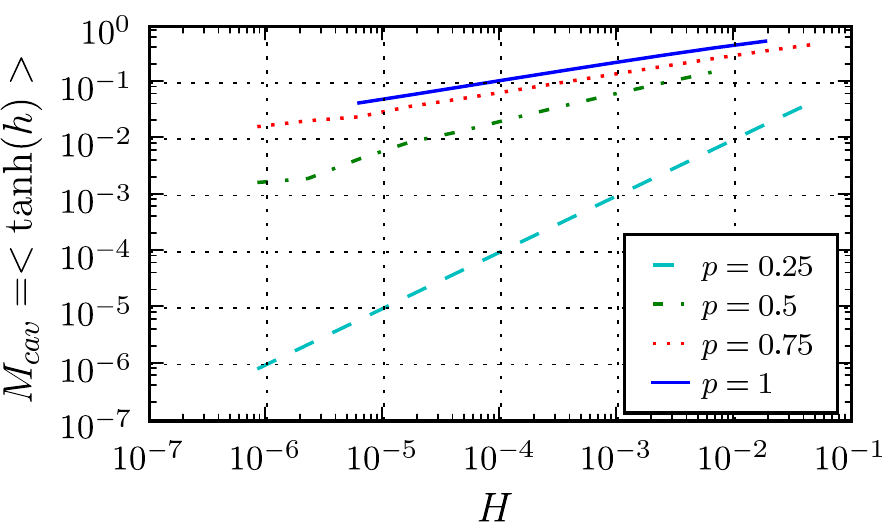}
  \caption{Critical magnetization in an external field along the
    $J=\infty$ line. The slope of the log-log curves
    indicates the critical exponent $\delta$ associated with each of
    the four points $p=0.25, 0.5, 0.75, 1$ along the phase
    boundary. We find $\delta=1$ for $p=0.25<p_c=0.5$, $\delta=1/2$
    for $p=0.5$ and $\delta=1/3$ for $p=0.75$ and $1$.}
  \label{fig:cav-critexp}
\end{figure}

Finally, Figure \ref{fig:cav-critexp} confirms numerically the
critical exponents derived using the moment flow analysis of Section
\ref{sec:preliminaries} and the exact solution of Section
\ref{sec:cluster}.

\bibliography{papers,griffiths}

\end{document}